\documentclass[12pt,preprint]{emulateapj}

\newcommand{\be}{\begin{equation}}
\newcommand{\ee}{\end{equation}}
\newcommand{\ba}{\begin{eqnarray}}
\newcommand{\ea}{\end{eqnarray}}

\begin{document}

\title{The extragalactic background light from the measurements of the attenuation of high-energy gamma-ray spectrum}

\author{Yan Gong$^1$ and Asantha Cooray$^1$}

\affil{$^1$Department of Physics \& Astronomy, University of California, Irvine, CA 92697}

\begin{abstract}

The attenuation of high-energy gamma-ray spectrum due to the electron-positron pair production against the extragalactic background light (EBL) provides an indirect method to measure the EBL of the universe. We use the measurements of the absorption features of the gamma-rays from blazars as seen by Fermi Gamma-ray Space Telescope to explore the EBL flux density and constrain the EBL spectrum, star formation rate density (SFRD) and photon escape fraction from galaxies out to $z=6$. Our results are basically consistent with the existing determinations of the quantities. We find a larger photon escape fraction at high redshifts, especially at $z=3$, compared to the result from the recent Ly$\alpha$ measurements. Our SFRD result is consistent with the data from both gamma-ray burst and UV observations in $1\sigma$ level. However, the average SFRD we obtain at $z\gtrsim 3$ matches the gamma-ray data better than the UV data. Thus our SFRD result at $z\gtrsim 6$ favors that it is sufficiently high enough to reionize the universe.

\end{abstract}


\maketitle

\section{Introduction}

The extragalactic background light (EBL) is the cumulative radiation from the stars and active galactic nuclei (ANGs) through the universe history. It is tightly related to the starlight from direct emission and dust re-radiation which are dominant in the ultraviolet (UV) to near-infrared (near-IR) and mid-IR to submillimeter bands, respectively \citep{Baldry03,Fukugita04,Franceschini08,Finke10, Stecker12, Inoue13}. Hence the measurements of EBL at different redshifts are important and fundamental for the understanding of the star formation history and galaxy formation and evolution. However, direct EBL observations is limited by large uncertainties due to the foreground contamination that needs to be accounted for (Hauser \& Dwek 2001). This makes the measurements of the EBL very difficult especially at high redshifts.

The high-energy gamma-rays interact with the EBL photons and generate the electron-positron pairs (Nikishov 1962; Gould \& Schreder 1966; Fazio \& Stecker 1970; Stecker et al. 1992). This effect can cause an absorption feature in the observed gamma-ray spectrum above a critical energy relative to its intrinsic spectrum. The evolution of the EBL therefore can be derived by the observations of the attenuated spectrum of the high-energy gamma-rays at different redshifts, which provides an indirect but feasible measurement of the EBL evolution. The difficulty of this method is how to determine the intrinsic spectrum of the gamma-ray sources and distinguish the absorption feature caused by the EBL from the intrinsic variation in the gamma-ray spectrum. 

Ackermann et al. (2012) reported the gamma-ray absorption feature in a sample of gamma-ray blazars in the redshift range $0.03<z<1.6$ using the Large Area Telescope (LAT) of the Fermi Gamma-ray Space Telescope\footnote{http://fermi.gsfc.nasa.gov/}. This sample contains 150 blazars of the BL Lacertae (BL Lac) type in the 3-500 GeV band, providing the constraints on the EBL spectrum from UV to the near-IR band. To determine the absorbed spectrum, they analyzed the sample in three independent redshift bins $z<0.2$, $0.2\le z<0.5$ and $0.5\le z<1.6$, and excluded the possibility that the attenuation is caused by the intrinsic properties of the gamma-ray sources.

We make use of this data set to constrain the EBL spectrum from UV to near-IR band, and extract the information of the photon escape fraction and the star formation history. Our EBL model is based on the work of Finke et al. (2010), which is dependent on a model for the stellar evolution, initial mass function (IMF), photon escape fraction, and the star formation rate density (SFRD). The Markov Chain Monte Carlo (MCMC) method is adopted in our constraint process, and we derive the average values and standard deviations for the EBL, photon escape fraction and SFRD from the obtained MCMC chains. We assume the flat $\Lambda$CDM with $\Omega_{\rm M}=0.27$, $\Omega_{\rm b}=0.046$ and $h=0.71$ for the calculation throughout the paper (Hinshaw et al. 2012).

\section{EBL Model}

We calculate the EBL spectrum based on the model proposed by Finke et al. (2010), where the EBL spectrum is evaluated by the initial mass function $\xi$($M_*$), the comoving star formation rate density $\dot{\rho}_*(z)$, and the photon escape fraction $f_{\rm esc}(\lambda,z)$ which denotes the fraction of photons that can escape a galaxy without absorption by interstellar dust. It gives an analytic relation between the EBL spectrum and the IMF, SFRD and $f_{\rm esc}$, and hence provides a good way to constrain these quantities using the EBL observational data.

We adopt the IMF model from Chabrier (2003) where it is expressed in two parts,
\ba
\xi(M_*) = \left\{\begin{array}{ll}
A\ {\rm exp}\left[ -\frac{({\rm log_{10}}M_*-{\rm log_{10}}M_*^c)^2}{2\sigma^2}\right] &, M_*\le 1 M_{\sun} \nonumber \\
B\ M_*^{-x} &, M_*>1 M_{\sun}
\end{array}\right.
\ea
where $M_*$ is the stellar mass, $A=0.158$, $M_*^c=0.079\ M_{\sun}$, $\sigma=0.69$, $B=4.43\times 10^{-2}$ and $x=1.3$. As this IMF model is consistent with the other models (e.g. Baldry \& Glazebrook 2003; Razzaque et al. 2009), we fix the values of these parameters when performing our MCMC fits.

We make use of the SFRD model proposed by Cole et al. (2001), which takes the form
\be \label{eq:SFRD}
\dot{\rho}_*(z) = \frac{a+bz}{1+(z/c)^d},
\ee
where $\dot{\rho}_*(z)$ is in $hM_{\sun}{\rm yr^{-1}Mpc^{-3}}$, and $a$, $b$, $c$ and $d$ are free parameters. At low redshifts with $z\lesssim2$, the SFRD can be constrained by the current observational data, and we take $a_{z\le2}=0.0118$, $b_{z\le2}=0.08$, $c_{z\le2}=3.3$ and $d_{z\le2}=5.2$ (Hopkins \& Beacom 2006, hereafter HB06)\footnote{These values are obtained using the IMF from Baldry \& Glazebrook (2003) which is well consistent with the IMF we use in Chabrier (2003).}. At $z>2$, the uncertainty of SFRD becomes very large because of uncertain dust extinction at the high redshifts. Hence we treat these four parameters as free and constrain them in our MCMC fits at $z>2$. Note that we don't consider the contribution from quiescent galaxies and AGNs in our model, since they are not the main components of the EBL which can contribute about 10\%$\sim$13\% emission to the total intensity (Dominguez et al. 2011).

The photon escape fraction $f_{\rm esc}$ should be a function of both the photon energy and redshift, and it is still not well determined by current observations. For simplicity, we assume $f_{\rm esc}$ is linearly increasing with the wavelength $\lambda$, which agree well with results from the observations of nearby galaxies (Driver et al. 2008). If the slope is independent of the redshift, then the photon escape fraction can be written as
\be \label{eq:fesc}
f_{\rm esc}(\lambda,z) = m(1+z)^n + p\ {\rm log_{10}}(\lambda/{\rm \mu m}).
\ee
Here $m$, $n$ and $p$ are free parameters and are needed as free parameter to fit in the MCMC process. When the photon energy is greater than $13.6$ eV, we set $f_{\rm esc}=0$, since most of the photons in this energy range would be absorbed by the neutral hydrogen in the galaxies.

\begin{figure*}[htbp]
\centerline{
\includegraphics[scale = 0.4]{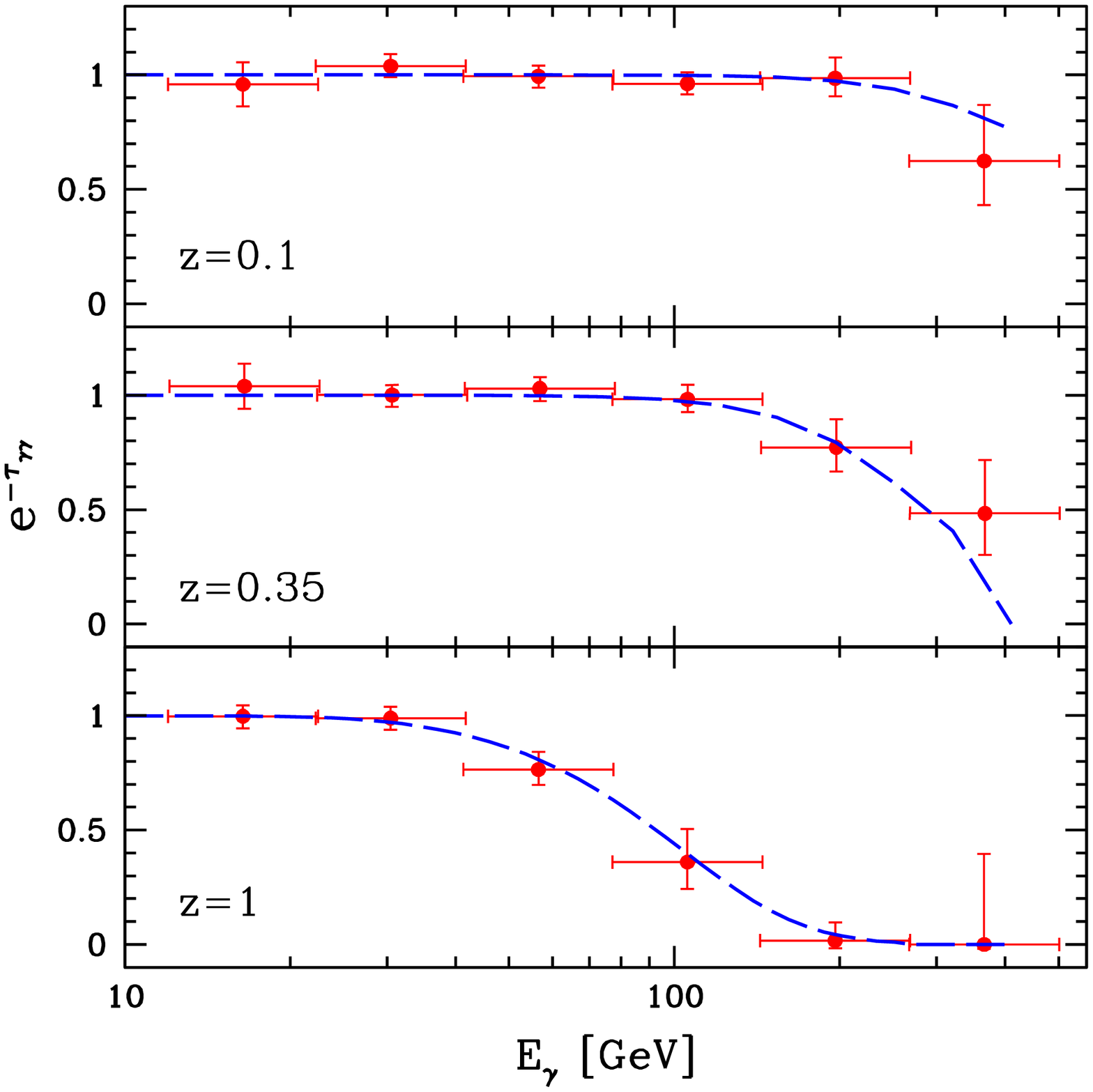}
\includegraphics[scale = 0.4]{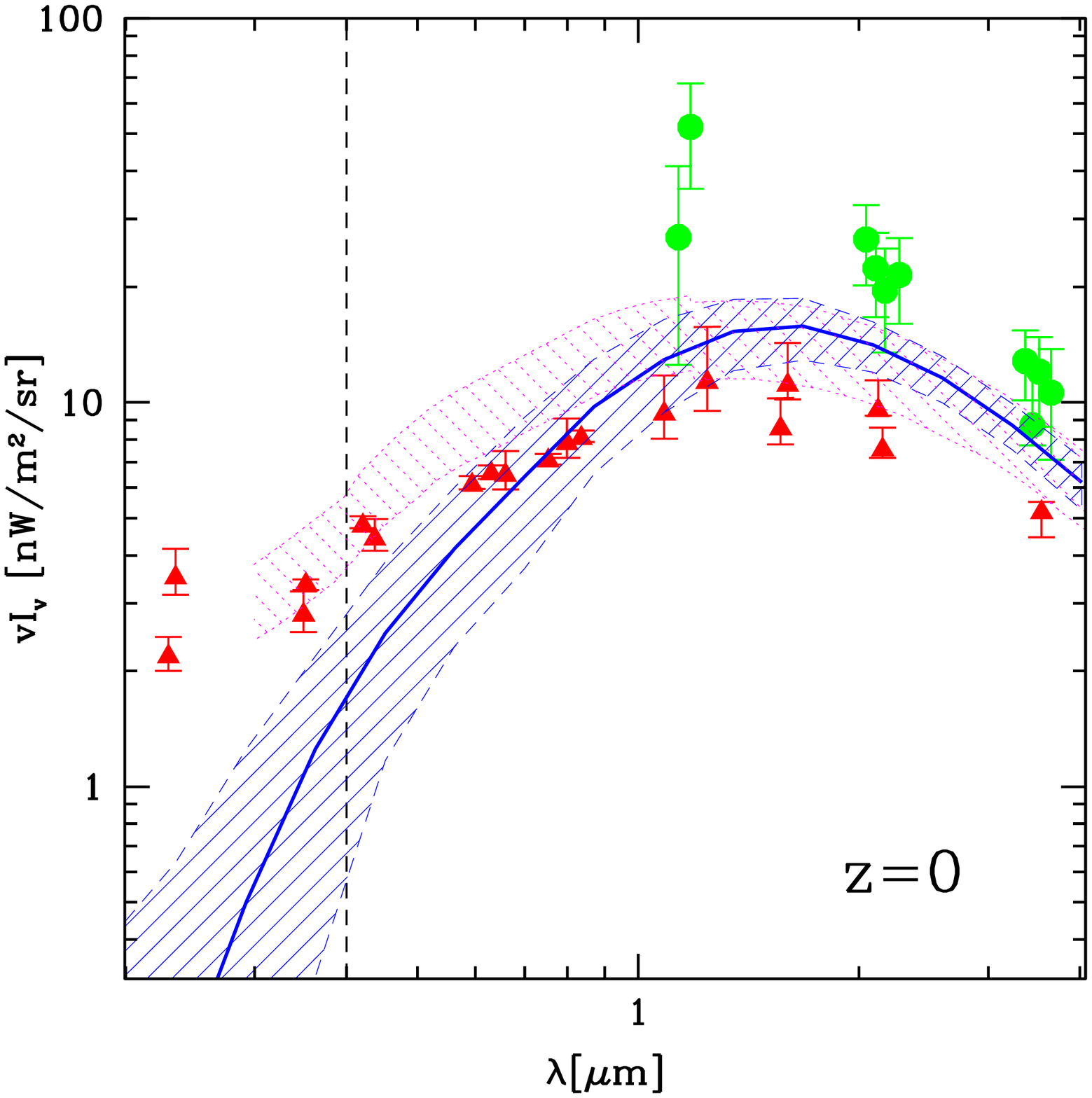} 
}
\caption{\label{fig:data_fit}  
{\it Left}: The data points of attenuation of gamma-ray spectrum by the EBL at three redshift bins used in this work and the best fits of our model (blue dashed lines). The redshift shown in each panel indicates the central redshift of the bins $z<0.2$, $0.2<z<0.5$ and $0.5<z<1.6$.
{\it Right:} The EBL spectrum at $z=0$ derived from the MCMC chains. We calculate the EBL at different wavelengths using each chain point, and find the the average values and the standard deviations ($1\sigma$) which are shown in blue solid line and shaded region respectively. For comparison, we also shows the data derived from the galaxy counts (red triangles) and direct measurements (green circles) at $z=0$ (Gilmore et al. 2012; Abramowski et al. 2013). The purple region is the 1$\sigma$ statistical contour estimated from several energy ranges of H.E.S.S. (Abramowski et al. 2013). Note that the EBL spectrum derived from our MCMC chains can only be constrained by gamma-ray attenuation data with $E_{\gamma} \gtrsim 200$ GeV, so it is not well consistent with the EBL data for $\lambda\lesssim 0.4\ \rm \mu m$ (dashed vertical line).}
\end{figure*}

Next we can estimate the comoving luminosity density $\epsilon\, j_{\epsilon}$ (in units of W/Mpc$^3$) where $\epsilon=h\nu/m_ec^2$ is the dimensionless photon energy. In our model, we consider both of the emission from stars and the re-radiation from interstellar dust, i.e. we have $\epsilon\, j_{\epsilon}=\epsilon\, j_{\epsilon}^{\rm star}+\epsilon\, j_{\epsilon}^{\rm dust}$. The comoving luminosity density from starlight at redshift $z$ is given by
\ba
\epsilon\, j_{\epsilon}^{\rm star}(\epsilon,z) &=& m_ec^2\epsilon^2\int^{M_*^{\rm max}}_{M_*^{\rm min}}{\rm d}M_*\ \xi(M_*) \\
&\times& \int_z^{z_{\rm max}} {\rm d}z' \left| \frac{{\rm d}t}{{\rm d}z'}\right| f_{\rm esc}(\epsilon,z') \dot{\rho}_*(z')\dot{N}_*(\epsilon, M_*, t_*).\nonumber
\ea
Here we take $M_*^{\rm min}=0.1\ M_{\sun}$, $M_*^{\rm max}=100\ M_{\sun}$, $z_{\rm max}=6$, and $\dot{N}_*(\epsilon, M_*, t_*)=\pi R_*^2\ c\ n_*(\epsilon,T_*)$ is the number of photons emitted per unit time per unit energy from a star with radius $R_*$ at lifetime $t_*$. The $n_*(\epsilon,T_*)$ is the stellar photon emission spectrum where $T_*$ is the stellar temperature. To estimate these quantities, i.e. $R_*$, $T_*$ and $n_*$, we assume the Planck spectrum for the starlight. This approximation is in a good agreement with the results given by Bruzual \& Charlot (2003) for stars with solar metallicity between 1 and 10 Gyr of age, which dominate the emission between 0.1 and 10 $\rm \mu m$ (Finke et al. 2010). We use a model of the Hertzsprung-Russell diagram to take into account different stellar evolution phases from the main-sequence to post main-sequence (e.g. the Hertzsprung gap, the giant branch, the horizontal branch, the asymptotic giant branch and the white dwarf) which take a stellar mass $0.1\le M_* \le 100$ $M_{\sun}$ (see Eggleton et al. 1989 for details). We also assume all stars have the solar metallicity, and it is constant over the cosmic history and stellar mass (Finke et al. 2010).

The radiation of dust which dominates the mid- and far-infrared bands is not important in our analysis here, since the gamma-ray sample we use is in the 3-500 GeV band. The process of photon-photon interaction between these gamma-rays and EBL photons would mainly occur in the near-infrared or higher energy EBL bands where the EBL photons are emitted directly from stars. Here we take the same dust emission model with three dust components used by Eq.~(11) of Finke et al. (2010). The proper EBL spectrum (or EBL intensity, in units of $\rm nW\, m^{-2}\, sr^{-1}$) at energy $\epsilon_p$ and redshift $z$ can be derived by
\be
\epsilon_p\, I(\epsilon_p,z) = (1+z)^4\, \frac{c}{4\pi} \int_{z}^{z_{\rm max}}{\rm d}z'\left|\frac{{\rm d}t}{{\rm d}z'}\right| \frac{\epsilon'\,j_{\epsilon}(\epsilon',z')}{1+z'},
\ee
where $\epsilon_p=(1+z)/(1+z')\ \epsilon'$ is the proper dimensionless photon energy at $z$. Also, it is easy to get the proper EBL energy density if we notice $\rho_b=(4\pi/c)\ \epsilon_p\,I$ which is in units of erg/cm$^3$. 

After obtaining the EBL spectrum, we can further estimate the optical depth for gamma-ray absorption with observed energy $E_{\gamma}$ emitted at redshift $z_0$ 
\ba \label{eq:tau}
\tau_{\gamma \gamma}(E_{\gamma},z_0) &=& \int_0^{z_0} {\rm d}z' \frac{{\rm d}l}{{\rm d}z'} \int_{-1}^1 {\rm d}u \frac{1-u}{2} \\
&\times& \int_{E_{\rm min}}^\infty {\rm d}E_{\rm b}\, n_{\rm b}(E_{\rm b},z') \sigma_{\gamma \gamma}(E_{\gamma}(1+z'), E_{\rm b}, u) \nonumber,
\ea
where ${\rm d}l/{\rm d}z$ is the cosmological line element, $E_{\rm b}=\epsilon_p\times m_ec^2$ is the proper photon energy of the EBL background at z, $u={\rm cos}(\theta)$ where $\theta$ is the angle of incidence, and $\sigma_{\gamma \gamma}$ is the cross-section of the pair production. The $n_{\rm b}(E_{\rm b},z)=\rho_b/E_{\rm b}$ is the proper number density of EBL photons at z. The $E_{\rm min}$ is the minimum threshold energy of EBL photons that can interact with a gamma-ray of observed energy $E_{\gamma}$
\be \label{eq:Emin}
E_{\rm min} = \frac{2m_{\rm e}^2c^4}{E_{\gamma}(1+z)(1-u)},
\ee
where $m_{\rm e}$ is the mass of electron. Then the intrinsic gamma-ray spectrum is modified to be
\be
F_{\gamma}^{\rm obs}(E_{\gamma}) = F_{\gamma}^{\rm int}(E_{\gamma})\ e^{-\tau_{\gamma \gamma}}.
\ee
Therefore, we can use the observations of attenuation of gamma-ray spectrum to compare with our theoretical model and constrain the free parameters that describe the SFRD and $f_{\rm esc}$ in the model.

\section{MCMC fitting}

We employ the MCMC method to perform the constraints. There are several advantages for this method, and the most important one is the time cost of the computations linearly increase with the number of the free parameters. Thus this method is suitable to fit a large set of parameters in an acceptable computation time. The Metropolis-Hastings algorithm is used in the MCMC process to decide possibility of accepting a new chain point (Metropolis et al. 1953; Hastings 1970). We use a Gaussian sampler with adaptive step size to estimate the proposal density matrix (Doran \& Muller 2004), and assume the uniform prior probability distribution for the parameters.

The $\chi^2$ distribution is employed to calculate the likelyhood function which is given by
\be
\chi^2 = \sum_{i=1}^N \frac{\left[ {\rm exp}(-\tau_{\gamma \gamma}^{\rm obs})-{\rm exp}(-\tau_{\gamma \gamma}^{\rm th}) \right]^2}{\sigma_i^2},
\ee
where N is the number of the data, $\tau_{\gamma \gamma}^{\rm th}$ is the theoretical optical depth given by Eq.~(\ref{eq:tau}), and $\tau_{\gamma \gamma}^{\rm obs}$ and $\sigma_i$ are the optical depth and errors from the observations. 

We use the observational data in Ackermann et al. (2012), where they provide the measurements of the absorption feature derived from 150 blazars in the 3-500 GeV band of the Fermi-LAT survey (see the data points in the left panel of Fig.~\ref{fig:data_fit}). This sample covers the redshift range from 0.03 to 1.6, and gives the absorption feature in three redshift bins with the central redshifts $z_{\rm c}\simeq$ 0.1, 0.35 and 1, respectively. We finally have 18 data points (6 in each redshift range), and we fit them all in three redshift bins with our model.

We have seven free parameters in our model: $a$, $b$, $c$ and $d$ in the SFRD given in Eq.~(\ref{eq:SFRD}), and $m$, $n$ and $p$ in the $f_{\rm esc}$ shown in Eq.~(\ref{eq:fesc}). As we assume uniform prior for the parameters in the MCMC process, their ranges are set as $a\in(0,0.1)$, $b\in(0,1)$, $c\in(0,6)$, $d\in(0,10)$, $m\in(-4,4)$, $n\in(-2,2)$ and $p\in(0,3)$. These ranges are chosen according to the relevant models (Hoplins \& Beacom 2006; Driver et al. 2008) and empirical experience. We perform some pre-runs to check these ranges and make sure that they have the correct physical meaning and there is no the other maximum out of these ranges. Note that we fix $a$, $b$, $c$ and $d$ to be the values in the HB06 model when $z\le2$, since the SFRD is relatively well-determined in this redshift range by the current observations as we discuss in the last section.

We run 8 parallel chains and obtain about $10^5$ points sampled in each chain after the convergence determined by the criterion of Gelman \& Rubin (1992). After the burn-in process and thinning the chains, we merge them into one chain and finally collect about 10000 points that are used to investigate the probability distributions of the parameters and statistical quantities of the components in the model. More details of our MCMC method can be found in Gong \& Chen (2007).

\section{Results}

In the left panel of Fig.~\ref{fig:data_fit}, we show the data of the attenuation of the gamma-ray spectrum by the EBL background in the three redshift bins, and the best fits of our model are denoted in red dashed lines. We fit 18 data points in the three redshift bins simultaneously and perform a global fitting for all seven free parameters in our model. The best-fits and 1$\sigma$ errors (68.3\% confidence level) of the parameters derived from the 1-D probability distribution function (PDF) are $a=0.055^{+0.041}_{-0.050}$, $b=0.57^{+0.43}_{-0.54}$, $c=3.9^{+2.0}_{-3.3}$, $d=4.0^{+5.5}_{-3.8}$, $m=0.32^{+0.18}_{-0.11}$, $n=1.4^{+0.4}_{-0.5}$ and $p=2.2^{+0.8}_{-1.4}$. The data are measured in three redshift bins $z<0.2$, $0.2<z<0.5$ and $0.5<z<1.6$. We take central redshifts of $z=0.1$, $z=0.35$ and $z=1$ to perform our theoretical calculation. Note that our fitting results are only from the UV to near-IR bands of the EBL out to $1$ $\rm \mu m$ at $z=0$.

The EBL spectrum at $z=0$ derived from our MCMC chains are shown in the right panel of Fig.~\ref{fig:data_fit}. We calculate the EBL flux density for each chain point (which is a 7-D point and contains the values of seven free parameters in our model) at different wavelengths, and estimate the average values (blue solid line) and standard deviations ($1\sigma$, blue region). The EBL data evaluated from the galaxy counts (red triangles) and direct measurements (green circles) are also shown for comparison (Gilmore et al. 2012; Abramowski et al. 2013). The purple region gives the 1$\sigma$ statistical contour derived from different energy bands of High Energy Stereoscopic System (H.E.S.S.) (Abramowski et al. 2013). We find the EBL spectrum from our MCMC chains are consistent with these observational results at $\lambda \gtrsim 0.4\ \rm \mu m$ (vertical dashed line). For $\lambda \lesssim 0.4\ \rm \mu m$, the energy of gamma-ray which can interact with the EBL photons is less than 200 Gev (see Eq.~(\ref{eq:Emin})), and the data points of optical depth $\tau_{\gamma \gamma}$ are close to 0 as shown in the left panel, which can not give stringent constraints on the EBL at $z=0$ in this regime.

\begin{figure}[htb]
\includegraphics[scale = 0.4]{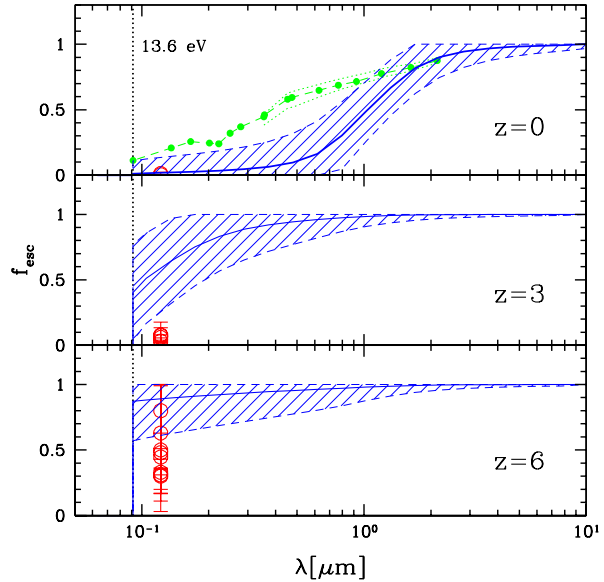}
\caption{\label{fig:fesc} The photon escape fraction as a function of $\lambda$ at $z=0$, 3 and 6 derived from the MCMC chains. The average values and standard deviations ($1\sigma$) are shown in blue solid line and shaded region respectively. The vertical black dotted line indicates the wavelength of the hydrogen ionization energy at 13.6 eV ($\sim$ 912 $\rm \AA$), and we set $f_{\rm esc}=0$ in the calculation when the wavelength is less than that. The red circles with error bars denote the data of Ly$\alpha$ escape fraction measurements around these three redshifts given by Hayes et al. (2011) and Blanc et al. (2011) (and references therein). The green points and lines in the upper panel give the results and errors from nearby galaxies (Driver et al. 2008).
}
\end{figure}

In Fig.~\ref{fig:fesc}, we show the $f_{\rm esc}(\lambda,z)$ at three redshifts $z=0$, $z=3$ and $z=6$. The $f_{\rm esc}$ are calculated by each MCMC chain point at different wavelengths and redshifts, and the average values and standard deviations ($1\sigma$) are shown in blue solid line and shaded region respectively. The vertical black dotted line denotes the hydrogen ionization energy at 13.6 eV ($\sim$ 912 $\rm \AA$), and note that the $f_{\rm esc}$ is set to be 0 when the photon energy is greater than 13.6 eV in our calculation since most of these photons would be absorbed by the neutral hydrogen gas in galaxies. The red circles with error bars are the Ly$\alpha$ escape fraction derived from the Ly$\alpha$ luminosity function around these three redshifts(Hayes et al. 2011, Blanc et al. 2011 and references therein). We find our $f_{\rm esc}(\lambda,z)$ results well agree with these results at $z=0$ and 6, but are higher at $z=3$. The green points and lines in the upper panel are the average photon escape fraction and errors derived from 10,000 nearby galaxies (Driver et al. 2008). Our result is lower than theirs, especially around $\lambda=0.5\ \rm \mu m$, but consistent at longer wavelengths with $\lambda\gtrsim 1\ \rm \mu m$.

\begin{figure}[htb]
\includegraphics[scale = 0.4]{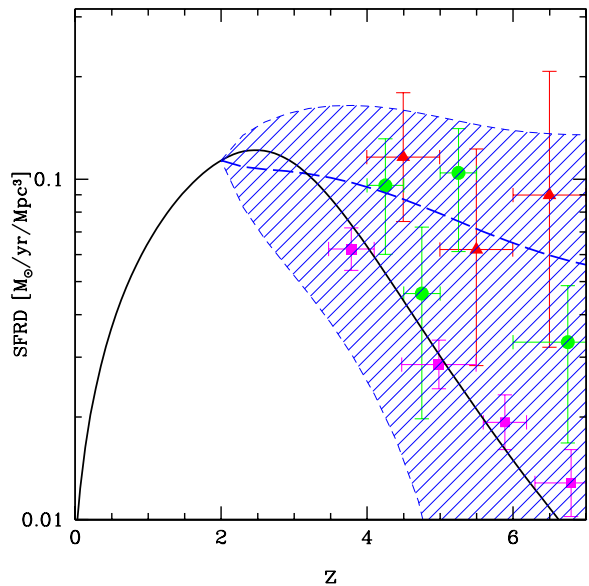}
\caption{\label{fig:SFRD} The star formation rate density derived from our MCMC chains. The black solid line shows the model obtained by Hopkins \& Beacom (2006), and the blue dashed line and the shaded region indicate the average values and standard deviations ($1\sigma$) estimated by the MCMC chains. The red triangles and green circles are the data given by GRB measurements from Kistler et al. (2009) and Robertson \& Ellis (2012), respectively. The purple squares are the data obtained by integrating UV luminosity functions shown in Bouwens et al. (2012).
}
\end{figure}

In Fig.~\ref{fig:SFRD}, we show the SFRD at different redshifts. The blue dashed line and the shaded region are the average SFRD values and standard deviations ($1\sigma$) evaluated by the MCMC chains, respectively. The black solid line denotes the model from Hopkins \& Beacom (2006) with the IMF of Baldry \& Glazebrook (2003). As addressed in Section 2, we fix SFRD to be the HB06 model at $z\le 2$ in our MCMC process, since it can be well constrained by the current observations. We find the average SFRD (blue dashed line) from our MCMC chains is higher than the HB06 model at $z\gtrsim3$ and the data from the UV observations (Bouwens et al. 2012) with flatter slope, but it agrees with the data from the gamma-ray burst measurements shown in red triangles (Kistler et al. 2009) and green circles (Robertson \& Ellis 2012, ``low-Z SFR" model). Also, we should note that our result is consistent with both of the gamma-ray and UV data in 1$\sigma$ level given the large uncertainty. This implies our SFRD result favors that it is alone sufficient to reionize the universe (Madau et al. 1999). 

\section{Discussion and conclusion}

We explore the EBL spectrum at near-IR to UV bands by fitting the gamma-ray attenuation data detected by Fermi-LAT measurements shown in Ackermann et al. (2012). Our EBL model is based on earlier work of Finke et al. (2010). This model can fit the gamma-ray attenuation data well in all three redshift bins within $0.03\le z\le1.6$. After obtaining the MCMC chains, we derive the average values and standard deviations of the EBL spectrum $\nu I_{\nu}(\lambda,z=0)$, $f_{\rm esc}(\lambda,z)$ and $\dot{\rho}_*(z)$ respectively from chain points. Also, we compare our results with the corresponding observational data, and find they are basically consistent with these observations in the regime of the gamma-ray attenuation data used in the constraints. 

The $f_{\rm esc}$ we get agrees with the data of Ly$\alpha$ escape fraction measurements at $z=0$ and 6, but a bit larger at $z=3$. Also, it is smaller than the result of Driver et al. (2008) around $\lambda=0.5\ \rm \mu m$ at $z=0$. For the star formation history, we obtain a higher average SFRD (blue dashed line in Fig.~\ref{fig:SFRD}) at $z\gtrsim 3$ with flatter slope than the result from the HB06 model and UV data. But note that our results in fact still consistent with theirs in $1\sigma$ level given the large constraint uncertainty. On the other hand, our average SFRD matches the results given by GRB measurements very well, and this indicates that our SFRD has a trend to favor that the star formation alone at high redshifts could reionize the universe.

Recently, Orr et al. (2011) and Yuan et al. (2012) also perform constraints on the EBL spectrum using the gamma-ray observations from Fermi and ground-based air Cherenkov telescopes with 12 and 7 blazars respectively. Their results around near-IR band are consistent with ours in 1$\sigma$ level but are higher than ours in the optical band. However, our EBL spectrum in the optical band agrees well with that from Dominguez et al. (2011) in which they use the observed galaxy luminosity function and galaxy SED-type fractions to derive the EBL spectrum.

\begin{acknowledgments}
This work was supported by NSF CAREER AST-0645427.
\end{acknowledgments}



\begin{thebibliography}{61}

\bibitem[Abramowski et al.(2013)]{Abramowski13}
Abramowski, A., et al. 2013, A\&A, 550, 4

\bibitem[Ackermann et al.(2012)]{Ackermann12}
Ackermann, M., et al. 2012, Science, 338, 1190


\bibitem[Baldry \& Glazebrook(2003)]{Baldry03}
Baldry, I.~K. \& Glazebrook, K. 2003, ApJ, 593, 258


\bibitem[Blanc et al.(2011)]{Blanc11}
Blanc, G.~A., Adams, J.~J., Gebhardt, K., et al. 2011, 736, 31


\bibitem[Bouwens et al.(2012)]{Bouwens12}
Bowwens, R.~J., Illingworth, G.~D., Oesch, P.~A., et al. 2012, ApJ, 754, 83


\bibitem[Bruzual \& Carlot(2003)]{Bruzual03}
Bruzual, G. \& Carlot, S. 2003, MNRAS, 344, 1000-1028


\bibitem[Chabrier(2003)]{Chabrier03}
Chabrier, G. 2003, Publ. Astron. Soc. Pac., 115, 763-796


\bibitem[Cole et al.(2001)]{Cole01}
Cole, S., et al. 2001, MNRAS, 326, 255

\bibitem[Dominguez et al.(2011)]{Dominguez11}
Dominguez, A., et al. 2011, MNRAS, 410, 2556


\bibitem[Doran \& Muller(2004)]{Doran04}
Doran, M. \& Muller, C.~M. 2004, J. Cosmol. Astropart. Phys., 09, 003


\bibitem[Driver et al.(2008)]{Driver08}
Driver, S.~P., Popescu, C.~C., Tuffs, R.~J., et al. 2008, ApJ, 678, L101


\bibitem[Eggleton et al.(1989)]{Eggleton89}
Eggleton, P.~P., Tout, C.~A. \& Fitchett, M.~J. 1989, ApJ, 347, 998


\bibitem[Fazio \& Stecker(1970)]{Fazio70}
Fazio, G.~G. \& Stecker, F.~W. 1970, Nature, 226, 135


\bibitem[Finke et al.(2010)]{Finke10}
Finke, J.~D., Razzaque, S. \& Dermer, C.~D. 2010, ApJ, 712, 238-249


\bibitem[Franceschini et al.(2008)]{Franceschini08}
Franceschini, A., Rodighiero, G. \& Vaccari, M. 2008, A\&A, 487, 837-852


\bibitem[Fukugita \& Peebles(2004)]{Fukugita04}
Fukugita, M. \& Peebles, P.~J.~E. 2004, ApJ, 616, 643


\bibitem[Gelman \& Rubin(1992)]{Gelman}
Gelman, A. \& Rubin, D. 1992, Stat. Sci., 7, 457


\bibitem[Gilmore et al.(2012)]{Gilmore12}
Gilmore, R.~C., Somerville, R.~S., Primack, J.~R. \& Dominguez, A. 2012, MNRAS, 422, 3189-3207


\bibitem[Gong \& Chen(2007)]{Gong07}
Gong, Y. \& Chen, X. 2007, Phys. Rev. D., 76, 123007

\bibitem[Gould \& Schreder(1966)]{Gould66}
Gould, R.~J. \& Schreder, G. 1966, Phys. Rev. Lett., 16, 252 


\bibitem[Hastings(1970)]{Hastings70}
Hastings, W. K. 1970, Biometrika, 57, 97


\bibitem[Hauser \& Dwek(2001)]{Hauser01}
Hauser, M.~G. \& Dwek, E. 2001, ARA\&A, 39, 249


\bibitem[Hayes et al.(2011)]{Hayes11}
Hayes, M., Schaerer, D., Goran, O., et al. 2011, ApJ, 730, 8


\bibitem[Hinshaw et al.(2012)]{Hinshaw12}
Hinshaw, G., Larson, D., Komatsu, E., et al. 2012, arXiv:1212.5226


\bibitem[Hopkins \& Beacom(2006)]{Hopkins06}
Hopkins, A.~M. \& Beacom, J.~F. 2006, ApJ, 651, 142


\bibitem[Inoue et al.(2013)]{Inoue13}
Inoue, Y., Inoue, S., Kobayashi, M.~A.~R., et al. 2013, ApJ, 768, 197


\bibitem[Kistler et al.(2009)]{Kistler09}
Kistler, M., Yuksel, H., Beacom, J.~F., Hopkins, A.~M. \& Wyithe, J.~S.~B. 2009, ApJ, 705, L104-L108




\bibitem[Madau et al.(1999)]{Madau99}
Madau, P., Haardt, F. \& Rees, M. J. 1999, ApJ, 514, 648


\bibitem[Metropolis et al.(1953)]{Metropolis53}
Metropolis, N., Rosenbluth, A. W., Rosenbluth, M. N., Teller, A. H. \& Teller, E. 1953, JCP, 21, 1087


\bibitem[Nikishov(1962)]{Nikishov62}
Nikishov, A.~I. 1962, Soviet Phys. JETP, 14, 393


\bibitem[Orr et al.(2011)]{Orr11}
Orr, M.~R., Krennrich, F. \& Dwek, E. 2011, ApJ, 733, 77


\bibitem[Razzaque et al.(2009)]{Razzaque09}
Razzaque, S., Dermer, C.~D. \& Finke, J.~D. 2009, ApJ, 697, 483


\bibitem[Robertson \& Ellis(2012)]{Robertson12}
Robertson, B.~E. \& Ellis, R.~S. 2012, ApJ, 744, 95


\bibitem[Stecker et al.(1992)]{Steker92}
Stecker, F.~W., de Jager, O.~C. \& Salamon, M.~H. 1992, ApJ, 390, L49




\bibitem[Stecker et al.(2012)]{Stecker12}
Stecker, F.~W., Malkan, M.~A. \& Scully, S.~T. 2012, ApJ, 761, 128


\bibitem[Yuan et al.(2012)]{yuan12}
Yuan, Q., Huang, H., Bi, X. \& Zhang, H. 2012, arXiv:1212.5866



\end{thebibliography}
\end{document}